\documentclass[aps,twocolumn,superscriptaddress,showkeys,showpacs]{revtex4}
\usepackage{amssymb}
\usepackage{amsfonts}
\usepackage{amsmath}
\usepackage{amsthm}
\usepackage{epsfig}
\usepackage{color}
\usepackage{subfigure}
\usepackage[normalem]{ulem}

\graphicspath{{./}{./Figures/}}

\newcommand{\mbf}{\mathbf}

\newcommand{\matr}[1]{{{\mbf{#1}}}}    
\renewcommand{\vec}[1]{{\mbf{#1}}}  
\newcommand{\m}[1]{\begin{pmatrix}#1\end{pmatrix}}

\begin{document}

\title{Delay controls chimera relay synchronization in multiplex networks}

\author{Jakub Sawicki}
\affiliation{Institut f{\"u}r Theoretische Physik, Technische Universit\"at Berlin, Hardenbergstra\ss{}e 36, 10623 Berlin, Germany}
\author{Iryna Omelchenko}
\affiliation{Institut f{\"u}r Theoretische Physik, Technische Universit\"at Berlin, Hardenbergstra\ss{}e 36, 10623 Berlin, Germany}
\author{Anna Zakharova}
\affiliation{Institut f{\"u}r Theoretische Physik, Technische Universit\"at Berlin, Hardenbergstra\ss{}e 36, 10623 Berlin, Germany}
\author{Eckehard Sch{\"o}ll}
\email[corresponding author: ]{schoell@physik.tu-berlin.de}
\affiliation{Institut f{\"u}r Theoretische Physik, Technische Universit\"at Berlin, Hardenbergstra\ss{}e 36, 10623 Berlin, Germany}

\date{\today}

\begin{abstract}
We study remote (or relay) synchronization in multilayer networks between parts of one layer and their counterparts in a second layer, where these two layers are not directly connected. A simple realization of such a system is a triplex network where a relay layer in the middle, which is generally not synchronized, acts as a transmitter between two outer layers. We establish time delay in the inter-layer coupling as a powerful tool to control various partial synchronization patterns, in particular chimera states, i.e., complex patterns of coexisting coherent and incoherent domains. We demonstrate that the three-layer structure of the network allows for synchronization of the coherent domains of chimera states in the first layer with their counterparts in the third layer, whereas the incoherent domains either remain desynchronized or synchronized. By varying the topology of the relay layer, we study its influence on the remote synchronization in the outer layers. As model dynamics we use the paradigmatic FitzHugh-Nagumo system.
\end{abstract}


\maketitle

Complex networks are ubiquitous in nature and technology, and the analysis of their nonlinear dynamics and synchronization properties gives insight into diverse real-world systems \cite{PIK01,STR01a,ALB02a,NEW03,BOC18}. Recently, research has focussed on multilayer networks, which provide a description of systems interconnected through different types of links. The interplay of intra-layer interaction with inter-layer coupling opens up a plethora of phenomena in different fields, e.g. \cite{BOC14,DE13,DE15,KIV14}. A prominent example for such structures are social networks which can be described as groups of people with different patterns of contacts or interactions between them \cite{GIR02}. Other relevant applications are communication, supply, and transportation networks, for instance power grids, subway networks, or airtraffic networks \cite{CAR13d}. In biology, multilayer networks represent for instance neurons in different areas of the brain or neurons connected either by a chemical link or by an electrical synapsis \cite{BEN16,BAT17}. A special case of multilayer networks are multiplex topologies, where each layer contains the same set of nodes, and only pairwise connections between corresponding nodes from neighbouring layers exist \cite{ZHA15a,MAK16,JAL16,GHO16a,LEY17a,AND17}. 

Relay (or remote) synchronization between layers which are not directly connected is an intriguing phenomenon, which extends previously known relay synchronization between single systems, e.g., chaotic lasers \cite{SOR13}. The synchronization of network layers, which interact via an intermediate (relay) layer, has recently provoked much interest \cite{LEY18}. The simplest realization of such a system is a triplex network where a relay layer in the middle acts as a transmitter between the two outer layers. Network symmetries play an essential role in remote synchronization, where pairs of nodes synchronize despite their large distances on the network graph \cite{NIC13,GAM13,ZHA17,ZHA17a}.

In networks of isolated layers different types of dynamics have been observed, depending on the nature of the individual nodes and the topology within the layer. Besides complete synchronization, cluster synchronization, or desynchronized chaotic dynamics, more complex spatio-temporal patterns can be observed. Chimera states are prominent example of such patterns, they combine spatially coexisting domains of coherence and incoherence \cite{KUR02a,ABR04,SHI04,MOT10,PAN15,SCH16b}. Initially found in nonlocally coupled rings of identical oscillators, chimera states have recently been observed in a variety of network models with different topologies \cite{BUS15,BAN16,GHO16,MAJ17,KAS17,BUK17,BUK18}, and realized experimentally~\cite{HAG12,TIN12,MAR13,LAR13,GAM14,WIC13,SCH14a,ROS14a}. Chimera states are reminiscent of partially synchronized patterns in brain dynamics, such as unihemispheric sleep \cite{RAT00,RAT16} and epileptic seizure \cite{JIR13,JIR14,ROT14,AND16,CHO18}. 

It is the purpose of the present letter to extend the notion of relay synchronization from completely synchronized states to partial synchronization patterns in the individual layers and study various scenarios of synchronization of chimera states in a three-layer multiplex network of FitzHugh-Nagumo oscillators. This model is a paradigmatic system widely used in neuroscience and electrical engineering. 
Our analysis shows that the three-layer structure of the network gives rise to partial or full synchronization of chimera states in the outer layers via the relay layer. Our focus is on the control of the chimera synchronization patterns by time delay in the inter-layer coupling. Varying the topology of the relay layer allows to establish its effect on the remote synchronization in the outer layers. Our results might have widespread applications, including encrypted communication and neuronal dynamics.

\begin{figure}
\includegraphics[width=1.0\linewidth]{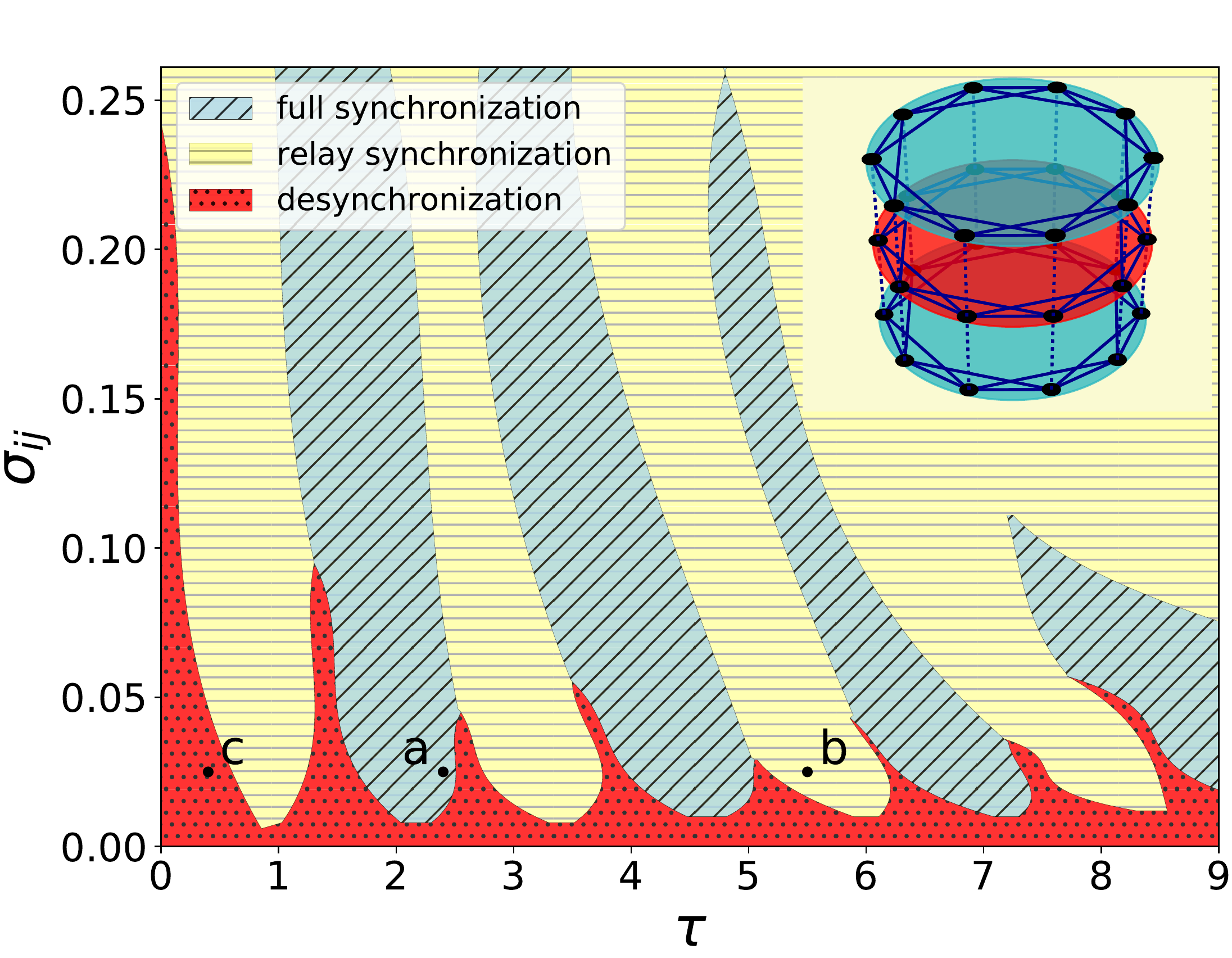}
\caption{(Color online) Relay synchronization tongues in the parameter plane of inter-layer coupling strength $\sigma_{ij}\equiv \sigma_{12}=\sigma_{23}$ and inter-layer coupling delay $\tau$: Inter-layer relay synchronization (horizontally hatched yellow region) occurs between regions of full inter-layer synchronization (diagonally hatched blue region) and desynchronized inter-layer dynamics (dotted dark red regions). Black dots (a, b, c) denote parameter values of the synchronization scenarios shown in Fig.\,\ref{fig:2}. Random initial conditions were used for all numerical simulations. Parameters: $\varepsilon=0.05$, $a=0.5$, $\sigma_i=0.2$, $R_i = 170$ for $i=1,2,3$, $\phi = \frac{\pi}{2}-0.1$, $N=500$. The inset shows a schematic triplex network. The middle layer $i=2$ (dark red) acts as relay layer between the two outer layers $i=1,3$ (light blue). }
\label{fig:1}
\end{figure}

The inset of Fig.\,\ref{fig:1} shows the configuration of a multiplex network with three layers (triplex). Each layer consists of a ring of $N$ identical FitzHugh-Nagumo (FHN) oscillators with non-local (intra-layer) coupling of coupling range $R_i$ in layer $i=1,2,3$, i.e., each oscillator is coupled with $R_i$ neighbors to the left and to the right. Layers $1$ and $3$ (light blue) are coupled through the intermediate layer $2$ (dark red), so that the middle layer acts as a relay between the two outer layers, but there is no inter-layer coupling between layers $1$ and $3$. The dynamical equations are given by
\begin{align}
\vec{\dot{x}}_k^i(t) =  \vec{F}(\vec{x}_k^i(t)) &+  \frac{\sigma_i}{2R_i} \sum^{k+R_i}_{l=k-R_i}\matr{H}[\vec{x}_l^i(t)-\vec{x}_k^i(t)] + \nonumber \\
&+ \sum^{3}_{j=1} \sigma_{ij} \matr{H}[\vec{x}_k^j(t-\tau)-\vec{x}_k^i(t)]
\label{eqn:gen1}
\end{align}
where $\vec{x}_k^i =(u, v)^T\in \mathbb{R}^2$, $i \in \{1,...,3\}$, $k \in \{1,...,N\}$ with all indices modulo $N$, denotes the set of activator ($u$) and inhibitor ($v$) variables, and the dynamics of each individual oscillator is governed by 
\begin{eqnarray}
\label{eq:localdyn}
\vec{F}(\vec{x})=
\left(\!
\begin{array}{*{1}{c}}
\varepsilon^{-1}(u-\frac{u^3}{3}-v)\\
u + a
\end{array}
\!\right),
\end{eqnarray}
where $\varepsilon > 0$ describes the time scale separation between fast activator and slow inhibitor, fixed at $\varepsilon = 0.05$ throughout this letter. Depending on the threshold parameter $a$ the single FHN elements exhibit either oscillatory ($|a|<1$) or excitable ($|a|>1$) behavior. Here we choose the oscillatory regime ($a=0.5$). The parameter~$\sigma_i$ denotes the intra-layer coupling strength, while $\sigma_{ij}$ is the inter-layer coupling strength. We use time delay $\tau$ only in the inter-layer coupling, since in real-world systems the transfer of information between two different layers is often slower than within one layer. 
In order to ensure constant row sum we choose the inter-layer coupling matrix as 
\begin{eqnarray}
\boldsymbol{\sigma}=\m{0 &  \sigma_{12} & 0 \\   \frac{\sigma_{12}}{2} & 0 &  \frac{\sigma_{23}}{2} \\  0 & \sigma_{23} & 0}
\end{eqnarray}
with $\sigma_{12}=\sigma_{23}$. The interaction is realized through diffusive coupling with coupling matrix 
\begin{eqnarray}
\matr{H}=\m{\varepsilon^{-1}\cos \phi& \varepsilon^{-1} \sin \phi\\ -\sin \phi&\cos \phi}
\end{eqnarray}
and coupling phase $\phi = \frac{\pi}{2}-0.1$. This coupling scheme, which consists predominantly of activator-inhibitor cross-coupling, is similar to a phase-lag of approximately $\pi/2$ in the Kuramoto phase oscillator model and has been chosen such that chimera states are most likely to occur~\cite{OME13}.
\begin{figure}
\includegraphics[width=1.05\linewidth]{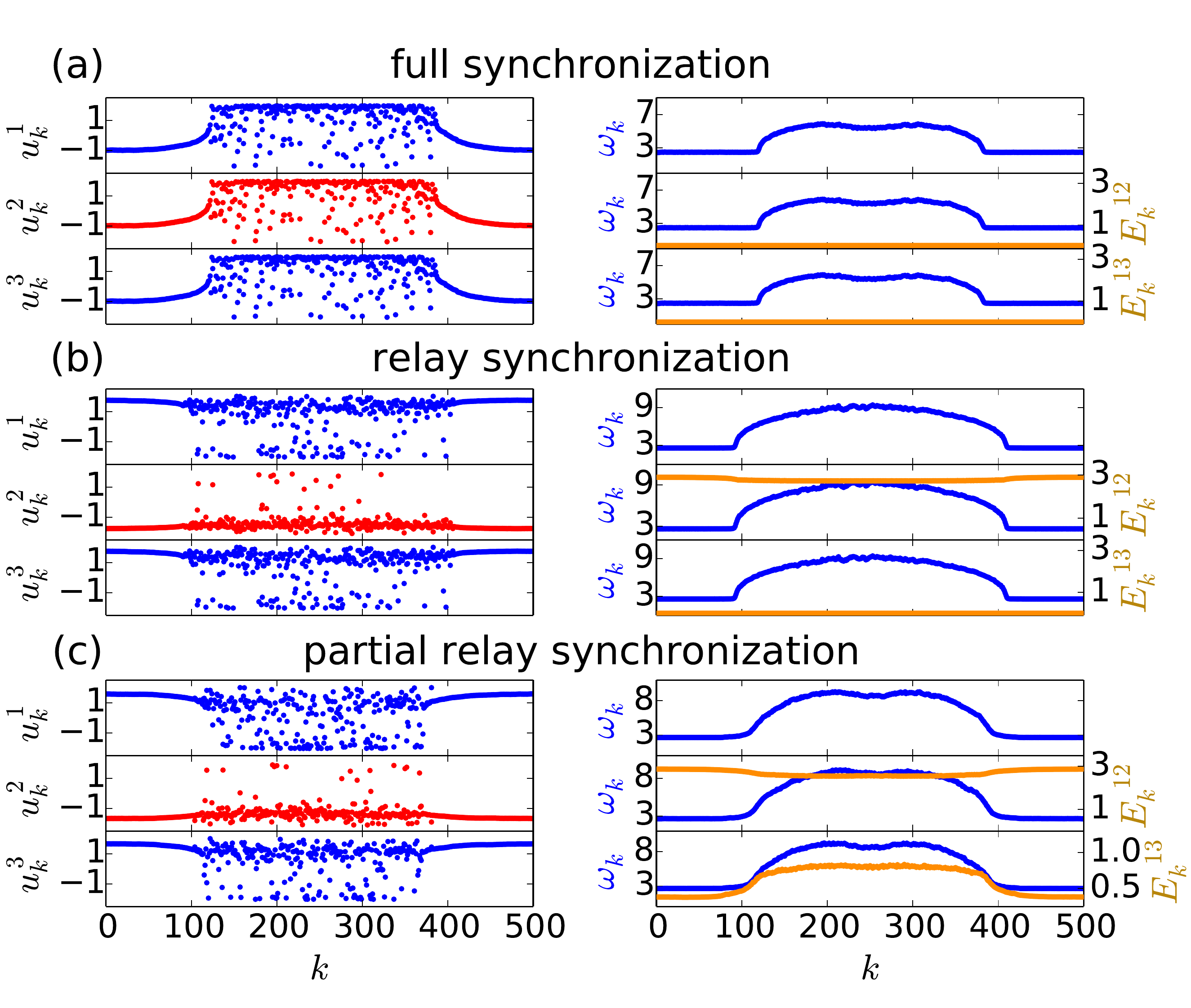}
\caption{Dynamics of the three layers for different values of delay time $\tau$, marked by black circles (a,b,c) in Fig.\,\ref{fig:1}: (a) full inter-layer synchronization for $\tau=2.4$. (b) relay inter-layer synchronization for $\tau=5.5$. (c) partial relay inter-layer synchronization between the outer layers for $\tau=0.4$. The left column shows snapshots of variables $u_k^i$ for all three layers $i=1,2,3$ (relay layer: red, outer layers: blue), whereas the right column shows the corresponding mean phase velocity profiles $\omega_k$ (dark blue) for each layer and inter-layer synchronization error $E^{ij}_k$ (orange). Inter-layer coupling is given by $\sigma_{ij}=0.025$, other parameters as in Fig.\,\ref{fig:1}.}
\label{fig:2}
\end{figure}
Generally, a time delay $\tau$ in the coupling often leads to spatially travelling patterns as shown in \cite{SAW17}. The same effect is observed for our multiplex network in case of delayed inter-layer coupling. Consequently, it is not possible to extract any information from measures calculated over a long time, e.g., the mean phase velocity profile and the local inter-layer synchronization error $E^{ij}_k$ introduced below. By detrending the data we can avoid this problem: After each time step in the numerical simulation  we re-index the nodes $k$ in such a way that $k' =(k + c)$, where $c$ is given by the center of the largest domain of the ring where for all $k$'s of that domain $\left \Vert \vec{x}_k(t) - \vec{x}_{k+1}(t) \right \Vert < \theta$ with a threshold $\theta$ chosen as $\theta = 0.25$~\cite{suppl}. 

For a single-layer network it is known that for appropriate coupling strength $\sigma_i$ and coupling range $R_i$ complex patterns of spatially coexisting coherent and incoherent dynamics, i.e., chimera states, can occur and they may be centered at different spatial locations depending on the initial conditions \cite{OME13}. On the other hand, it has been shown recently \cite{LEY18}, that in multiplex networks one can achieve synchronization of either neighboring or remote layers. However, the synchronization of complex spatio-temporal patterns like chimera states in multiplex networks is still largely unresolved. Here we establish the possibility to control partial synchronization patterns even of remote layers, in particular chimera states, by tuning the inter-layer coupling strength $\sigma_{ij}$ and delay $\tau$. Varying these two parameters allows for an overall control of the dynamical regimes in the network. 
An appropriate measure for synchronization between two layers $i,j$ is the global inter-layer synchronization error $E^{ij}$, defined by
\begin{eqnarray}
E^{ij}=\lim_{T \to \infty}\frac{1}{NT}\int_0^T \sum_{k=1}^N \left \Vert \vec{x}_k^j(t) - \vec{x}_k^i(t) \right \Vert \,dt,
\label{eq:Eij}
\end{eqnarray}
where $\left \Vert \cdot \right \Vert$ stands for the Euclidean norm, and the normalization by $N$ allows for better comparison of networks of different size~\cite{suppl}. 
First we consider three identical layers. Regarding the inter-layer synchronization three dynamical regimes are conceivable:
\begin{itemize}
\item {\it full inter-layer synchronization} where synchronization exists between all three layers ($E^{12}=E^{13}=0$)
\item {\it relay inter-layer synchronization} where synchronization exists just between the two outer layers ($E^{12}\neq0$ and $E^{13}=0$)
\item {\it inter-layer desynchronization} ($E^{12}\neq E^{13}\neq 0$)
\end{itemize}
Numerical simulations in Fig.\,\ref{fig:1} show that we can observe all scenarios depending on the parameters and the initial conditions (here: random initial conditions). When the layers are coupled weakly, they tend to behave independently (red dotted region): Each layer exhibits a chimera state but there is no synchronization between the layers. With increasing delay $\tau$ we observe a sequence of tongue-like regions in the parameter plane $(\tau, \sigma_{ij})$: Full inter-layer synchronization (blue regions with diagonal stripes) alternating with relay inter-layer synchronization (yellow regions with horizontal stripes). Exemplary snapshots of the dynamics in these synchronized regions are shown in Fig.\,\ref{fig:2}\,(a,b) (left column). We can observe full in-phase synchronization of all three layers for values of $\tau$ close to integer multiples of the period of the uncoupled system $T=2.3$, and relay inter-layer synchronization with anti-phase synchronization between the outer layers and the relay layer for half-integer multiples. Analytical calculations show that the period $T$ decreases with increasing $\sigma_{ij}$~\cite{suppl}. Therefore, due to the resonance condition of $\tau$ with respect to the intrinsic period $T$, the tongues are shifted to the left with increasing coupling strength $\sigma_{ij}$. The same effect occurs when $\tau$ equals higher multiples of the intrinsic period, where the tongues are shifted more strongly to the left and decrease in size, which is a general feature of resonance tongues in delay systems \cite{HOE05,YAN06}. To study the synchronization of chimera patterns between the layers in more detail, we use the local inter-layer synchronization error in dependence of each node $k$:
\begin{eqnarray}
E^{ij}_k=\lim_{T \to \infty}\frac{1}{T}\int_0^T  \left \Vert \vec{x}_k^j(t) - \vec{x}_k^i(t) \right \Vert \,dt.
\label{eq:Eijk}
\end{eqnarray}
This measure is useful in detecting those nodes which are synchronized between two layers, especially in the (red dotted) region of desynchronization in Fig.\,\ref{fig:1}. Exemplary dynamics inside this region are given in Fig.\,\ref{fig:2}\,(c): We can see the arc-shaped profiles for both mean phase velocity $\omega_k$ and local inter-layer synchronization error $E^{13}_k$. This means that the coherent parts of the chimera states are synchronized between the outer layer, whereas the incoherent parts are not. This kind of synchronization may be called {\it partial relay inter-layer synchronization} or {\it double chimera}, since it denotes coherence-incoherence behavior within the layers and between the layers.
It cannot be detected by the global inter-layer synchronization error $E^{ij}$, but the node-dependent local measure $E^{ij}_k$ gives us the possibility to distinguish this type of synchronization. In Figure \ref{fig:2} (right column) $E^{ij}_k$ is plotted (light orange) together with the mean phase velocity profile $\omega_k$ (dark blue) for a typical chimera state. The mean phase velocity of the oscillators is calculated as $\omega_k = 2\pi S_k / \Delta T$, $k=1,...,N,$ where $S_k$ denotes the number of complete rotations realized by the $k$th oscillator during the time $\Delta T$. Throughout the paper we use $\Delta T = 10000$.

\begin{figure}
\includegraphics[width=1.05\linewidth]{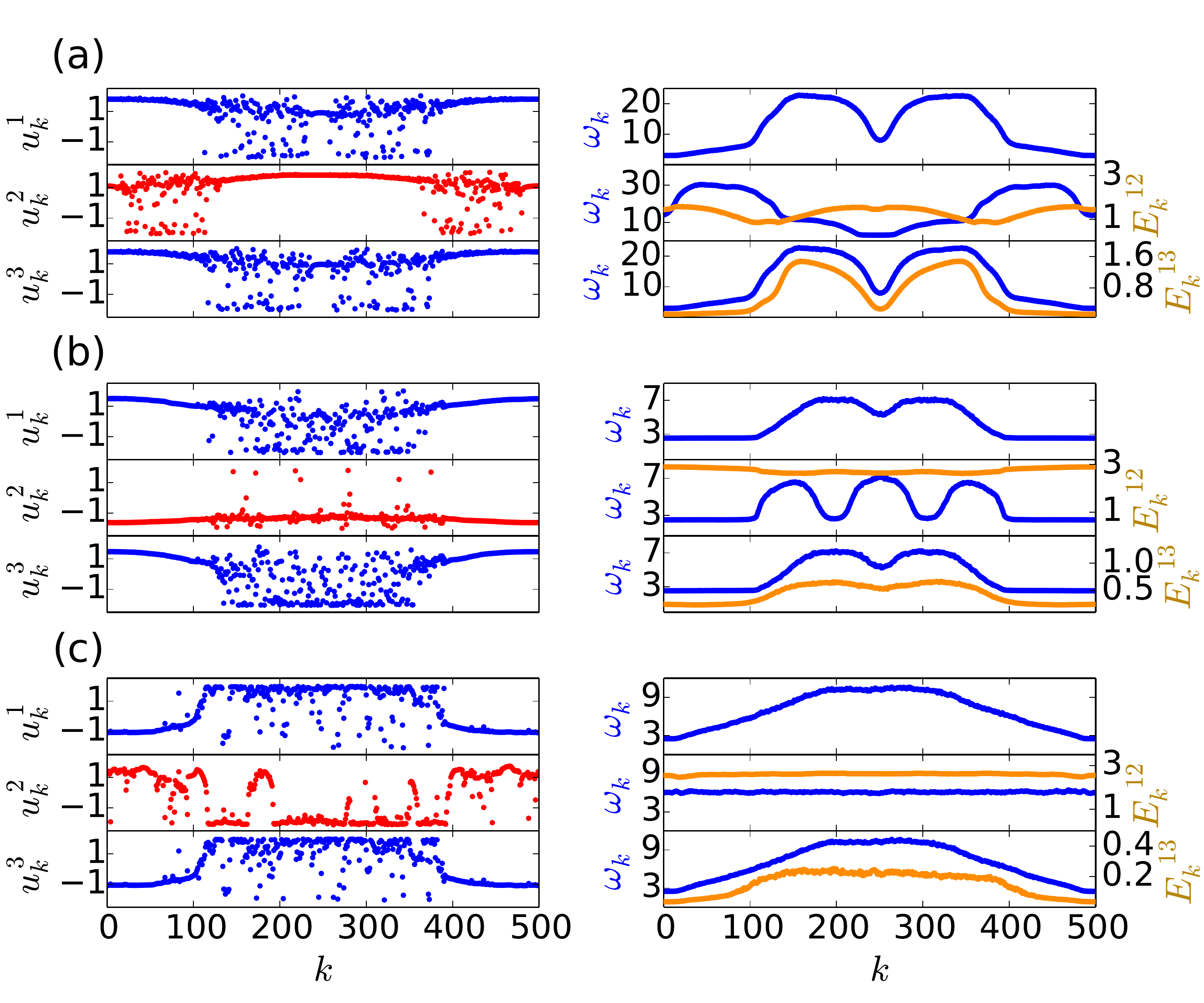}
\caption{Partial relay inter-layer synchronization between the outer layers: snapshots of variables $u_k^i$ (left column) for layers $i=1,2,3$ (relay layer: red, outer layers: blue), and mean phase velocity profiles $\omega_k$ (dark blue) and inter-layer synchronization error $E^{ij}_k$ (orange) in the right column. (a) $R_1=R_2=R_3=170$, $\tau=1.3$, $\sigma_{ij}=0.025$; (b) $R_1=R_3=150$, $R_2=130$, $\tau=0.4$, $\sigma_{ij}=0.015$; (c) $R_1=R_3=150$, $R_2=10$, $\tau=0.8$, $\sigma_{ij}=0.01$. Other parameters as in Fig.\,\ref{fig:1}.}
\label{fig:3}
\end{figure}

In our simulations we observe different intriguing types of partial relay inter-layer synchronization, for instance, Fig.\,\ref{fig:3}\,(a) depicts an example ($\tau=1.3$) where the relay layer exhibits anti-synchronization of chimera patterns: the coherent domain of the relay layer (red, middle panel) spatially coincides with the incoherent domains of the outer layers.

In addition, to demonstrate the robustness of our findings, we vary the topology of the relay layer compared to the outer layers by changing its coupling range.
Fig.\,\ref{fig:3}\,(b) shows partial relay inter-layer synchronization for the case of small mismatch of the coupling range in the relay and outer layers ($R_1 = R_3 = 150$, $R_2=130$). The middle layer exhibits a chimera state with three incoherent domains, in contrast to two in the outer layers, and the coherent domains in the relay layer and the outer layers are in anti-phase. The local synchronization error $E^{13}_k$ between the two outer layers is nonzero in the incoherent domains and vanishes for the coherent domains,
as a signature of partial relay synchronization.

Moreover, for large mismatch of the coupling ranges in the relay and outer layers, see Fig.\,\ref{fig:3}\,(c) where $R_1 = R_3 = 150$ and $R_2=10$, the relay layer is characterized by chaotic dynamics. This strongly chaotic dynamics of the relay naturally affects the chimera states in the outer layers, so that their mean phase velocity profiles (dark blue) are smeared out despite of detrending. Nevertheless, the coherent domains of the chimera states are synchronized between the outer layers, whereas the incoherent parts are not, as shown in the snapshots and the plot of $E^{13}_k$. Thus, the relay synchronization mechanism turns out to be robust with respect to changes of the relay layer topology. Preliminary studies show that this holds also for a mismatch of the excitation parameter $a$ between the layers~\cite{suppl}.

To conclude, we have shown that multilayer networks allow for intriguing remote synchronization scenarios. Relay synchronization of chimeras between the outer layers of a multiplex network is an example of such a scenario, where distant layers of the network synchronize in spite of the absence of direct connections between them. We have analyzed relay synchronization in a three-layer network of FitzHugh-Nagumo oscillators, with nonlocal coupling topologies within the layers, and have extended the notion of relay synchronization to chimera states. 

Chimera patterns can be observed in each network layer; they are usually strongly dependent on the initial conditions, and it is not possible to predict which part of the network will form coherent domains. By relay synchronization we can fix the location to the same position as in the other outer layer. Varying the strength of the coupling between the network layers, we observe various scenarios of synchronization of chimera states, either in all three layers, or only in the two outer layers. As measures we employ the global and local inter-layer synchronization errors and mean phase velocity profiles of the oscillators.

Time delay in the inter-layer coupling, which is ubiquitous in real-world systems, has been identified as a powerful tool for control of the patterns: It allows for observation of novel synchronization scenarios where the coherent domains of chimera states in the outer layers are synchronized, while the incoherent domains are not. The relay layer remains desynchronized and exhibits various multi-chimera patterns, or even chaotic dynamics. Furthermore, partial relay synchronization of chimeras states in the two outer layers has been realized in the form of intriguing double chimeras, where the coherent domains in both layers are synchronized, while the incoherent ones are not. By choosing an appropriate value for the time delay we can switch between the different synchronization scenarios.

Control of chimera patterns can also be effected by changing the topology in the intermediate layer. By varying the coupling range we find that even strongly diluted relay layers allow for remote synchronization of chimeras in the outer layers, while the relay layer stays in the chaotic regime.

We propose that our findings may be useful in the study of novel concepts for encrypted and secure communication, where relay synchronization of complex spatio-temporal patterns, for instance chimera states, can be employed. Since the dynamics of the intermediate (relay) layer is not synchronized, it does not transmit information to someone listening in. While relay synchronization of single chaotic lasers has been extensively investigated in the context of encrypted communication~\cite{SOR13}, here we have extended and generalized the concept of relay synchronization to multilayer networks, which exhibit much more complex dynamics.
As brain networks are often described as multilayer structures, our results may also help in elucidating complex scenarios of information processing in neural networks. Recent research in neuroscience indicates that many parts of the brain, e.g., thalamus, interneurons, and hippocampus, act as a relay that connects two different regions \cite{GUI02,WAN11g,VAN15,HAL17}. Our analysis of relay synchronization scenarios in multiplex networks could thus help to understand dynamical patterns in the human brain. 

This work was supported by DFG in the framework of Collaborative Research Center SFB~910.



\end{document}